\documentclass[a4paper]{VisionStyle}
\usepackage{epsfig}

\begin{document}

\title{Optical Counterparts of Ultraluminous X-Ray Sources}

\author{Manfred W. Pakull\inst{} \and Laurent Mirioni\inst{}
% \and   M.A.N.Y.\,Unknowns\inst{2} 
} 

\institute{
  Observatoire Astronomique de Strasbourg, UMR 7550, 
  11, rue de l'Universit\'e, F-67000 Strasbourg, France 
%\and 
%  A second institute, Somewhere, Sometown, Somecountry 
}

\maketitle 

\begin{abstract}
Despite much observational and theoretical effort little is presently known about
the nature of the luminous non-nuclear X-ray sources which appear to 
largely surpass the Eddington limit of a few solar masses. Here we present first
results of our OHP/ESO/CFHT optical survey of the environments of variable 
ultraluminous X-ray sources (ULX) in nearby galaxies. At the position of several
ULX we find emission nebul\ae\ of a few hundred parsecs diameter, and which
often show 
both low and high ionisation emission lines. The gas must therefore be either 
photoionized by hard XUV continua, or be shock-ionized in the expanding bubbles.
The nebul\ae\ have kinematic ages of some million years and appear to be directly 
linked to the highly energetic formation process of the compact ULX or being 
inflated by ongoing stellar wind/jet activity. The discovery of intense 
He~II$\lambda4686$ nebular recombination radiation together with
comparatively strong [O~I]\,$\lambda$6300 emission around the variable ULX in dwarf
galaxy Holmberg II has allowed us to show that the interstellar medium acually 
'sees' and reprocesses part of the $\sim 10^{40}$ erg/s measured at X-ray 
wavelengths, if we assume isotropic emission. Strong beaming into our line of sight
which has been advocated to avoid such high luminosities can thus be excluded, at 
least for this source. 

\end{abstract}

\section{Introduction}
\label{mwpakull-E3_sec:intro}
  
The luminosity function of compact accreting X-ray sour\-ces in the Local Group of 
Galaxies shows a well defined cut-off corresponding to the Eddington limit
$L_{\rm E}=1.3\times10^{38}$~M/M$_{\sun}$~erg/s for a few solar masses. Beyond
$L_{\rm E}$, stable accretion is no longer possible since radiation pressure 
would push off any material away from the compact star. 

Nevertheless, there is now growing evidence for a class of non-nuclear
X-ray sources in nearby galaxies which have individual luminosities higher than 
$10^{39}$ and up to some $10^{40}$~erg/s which, in parenthesis, is higher than the
total high energy output of the Local Group of Galaxies. Such luminosities 
imply compact stellar masses of at least several tens to some hundred M$_{\sun}$, 
if the Eddington limit is not to be violated; presently available stellar 
evolutionary models at solar metallicity on the other hand do not predict 
compact remnants that massive. 

These objects have become much advertised under the label 
of {\em intermediate-mass} black holes, i.e. 
BHs with masses in between the stellar and the supermassive AGN type variety 
(c.f. \cite*{mwpakull-E3:cm99}, and they are also variously referred to as 
SuperEddington sources, Inter\-medi\-ate-lu\-mi\-nosity X-ray Objects (IXO), 
and Ultraluminous X-ray sources (ULX). 
ULX are known to be a mixed bag of objects, 
including, apart from galactic foreground stars and background AGN, recent 
supernov\ae\ and very young SN remnants, to which the Eddington 
limit does of course not apply. However, many ULX - including several which have 
previously been thought to be SNR - are variable on short timescales 
of hours to days, suggesting a compact nature.
Particularly impressive examples of ULX have recently been revealed with the
{\it Chandra\/} observatory (e.g. \cite{mwpakull-E3:kaa01}, 
\cite{mwpakull-E3:bau01})
 
One possible solution to this puzzle has been to postulate highly anisotropic 
radiation in the form of 'beaming' towards the observer which would lessen 
the energy requirements, below $L_{\rm E}$ of a more conventional 
$\leq10M_{\sun}$ accretor (see, i.e. \cite{mwpakull-E3:king01}).

X-ray spectra of several ULX observed with ASCA have successfully been 
fitted with the "disc black body" model (often with an additional 
power-law component) assuming an optically thick accretion disk 
(\cite{mwpakull-E3:cm99}, \cite{mwpakull-E3:maka00}, \cite{mwpakull-E3:mkm01}). 
The model works rather well for the spectra of galactic BH 
candidates, and this has often been taken as supporting evidence for a 
massive accreting BH nature of ULX. However, in this picture problems
with, among others, the high inferred inner temperature parameter remain, which 
according to \cite*{mwpakull-E3:mkm01} might possibly
indicate the presence of either a rapidly spinning BH or the realization of
an opically thick advection-dominated accretion flow close to the BH.

Many of the ideas about the nature ULX will remain speculation unless we have 
supportive evidence from observations at other wavelengths. Moreover, a 
multi-wave\-length approach might help to understand formation and evolution of 
these interesting objects. We have therefore undertaken an 
observing program with the
aim to identify the optical counterparts of ULX and to study the local stellar
and interstellar population.

\section{Opical Follow up: Prospects}
\label{mwpakull-E3_sec:prosp}

ULX are only found outside the Local Group of Galaxies. This implies
that their optical counterparts are probably rather faint, even if they 
are of the optically brightest massive X-ray binary X-ray (MXRB) 
variety. An O star companion (for which we assume $M_{\rm V}\approx -5$)
would have $V>22$~mag at a 3 Mpc distance of the closest ULX and beyond. This
bleak perspective explains why not many optical follow-up observations of
ULX have been reported until recently. 

\begin{figure}
\centering
\includegraphics[width=\columnwidth]{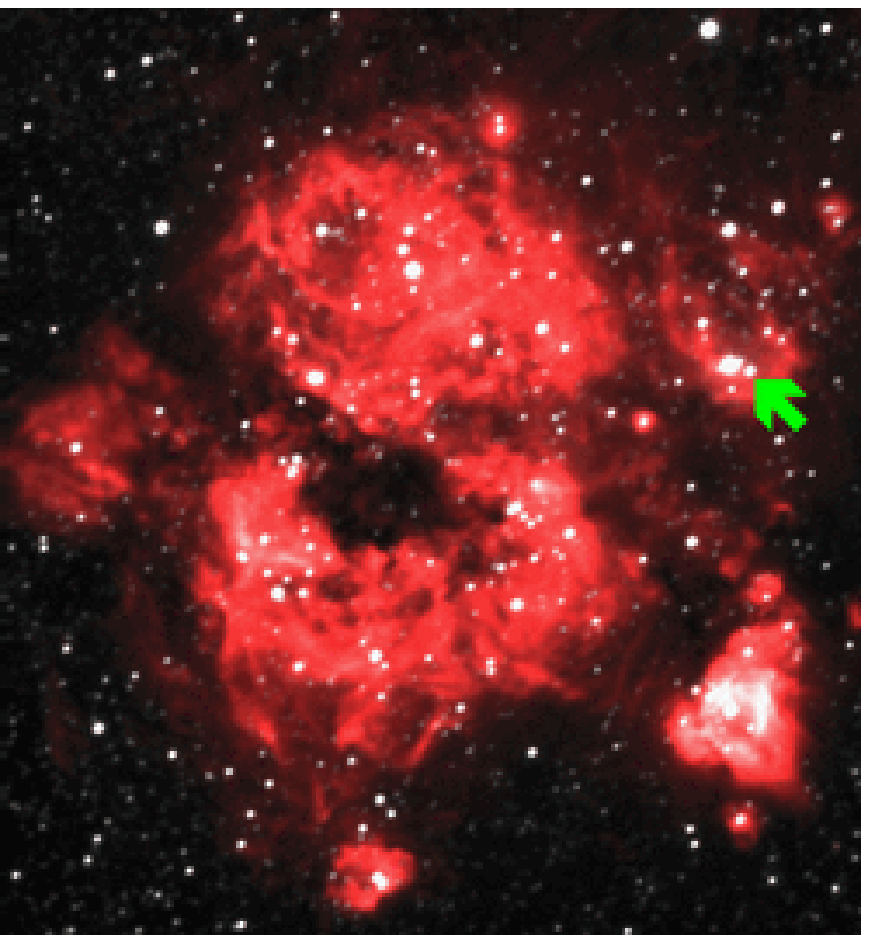}
%\vspace*{1cm}
\includegraphics[width=\columnwidth]{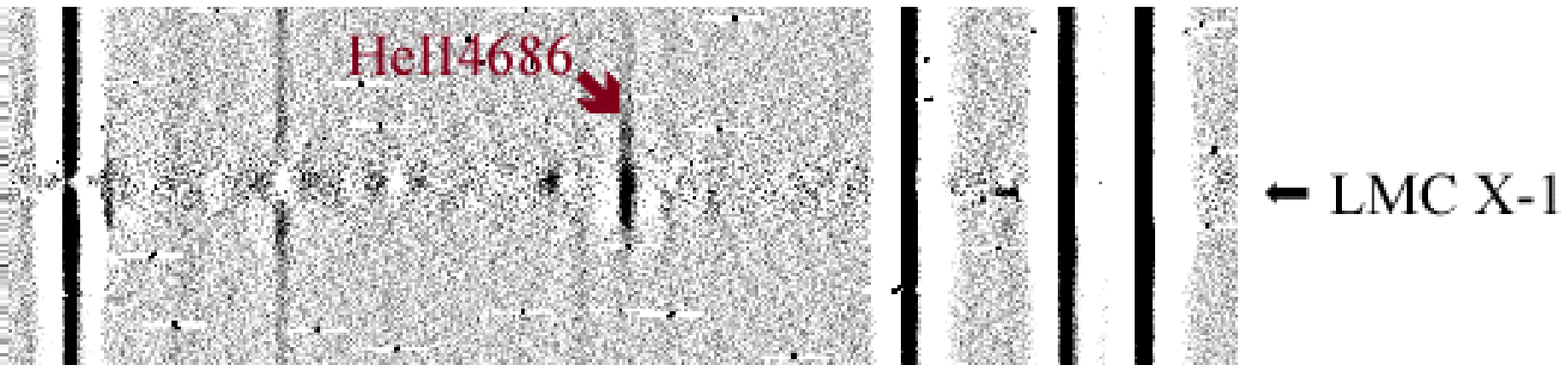}
\caption{The black hole candidate \object{LMC~X-1} and its X-ray 
ionized nebula N159F.  
H$_{\alpha}$ emission is shown in red. 
The O~7 main sequence type optical 
counterpart ($V\approx14.6$) is the fainter star of a close pair.  
The X-ray source is embedded in the $9\times13$ pc H~{\sc ii} region N~159F 
which is part of the larger complex \object{N159} in the Large Magellanic Cloud.
The lower image is a section of a long slit spectrum centered on the 
optical counterpart. It has been filtered in the spatial direction (effectively
removing the stellar continuum) in order to display the presence of the 
clearly extended narrow He\,II$\lambda$4686 emission in the nebula.
This emission is one of the hallmark of an X-ray ionized nebula (see text).  
The strong nebular lines are, from left to right, H$_{\gamma}$, H$_{\beta}$, 
and [O~III]$\lambda\lambda$4959,5007.}
\label{mwpakull-E3_fig:fig1}
\end{figure}

Potentially more spectacular are the possible interaction effects 
of X-ray sources with the diffuse interstellar medium.
The best-known example is the radio nebula
\object{W~50} (i.e. \cite{mwpakull-E3:dhgm98})
around the enigmatic \object{SS~433}. This nonthermal galactic nebula is 
a spherical shell with two lateral extensions, the "ears", that are
formed by interaction of the jets with the sourrounding gas. The 
$60\times120$ pc extent is that of a large supernova remnant which would
be easily resolvable at several Mpc distance. Another possibility 
is photoionization of interstellar gas by the X-ray source.
Until very recently, the only well-established X-ray ionized nebula (XIN -  
\cite{mwpakull-E3:pa86}) has been the Large Magellanic
Cloud nebula \object{N159F} in which the black hole candidate \object{LMC~X-1} is
embedded; see Figure~\ref{mwpakull-E3_fig:fig1}.

Here, more than fifteen years ago we detected extended 
narrow He\,II$\lambda4686$ recombination emission
which is exactly centered on the presumed O star counterpart. Since with the
exception of hot Wolf-Rayet stars with weak winds (\cite{mwpakull-E3:pa91}), no massive
star emits any significant amount of He$^+$ ionizing photons (h$\nu>\,$54 eV)
this high excitation line must be due to X-ray photoionization. In Sect. 
\ref{mwpakull-E3_sec:xin} we
will see that the luminosity of the He\,II$\lambda4686$ line even allows to estimate
the X-ray luminosity of the embedded source.

A further subject well worth further optical follow-up work is the
rather unexpected association of several optically selected supernova remnants
with ULX (\cite{mwpakull-E3:mi95}, \cite{mwpakull-E3:mf97},
\cite{mwpakull-E3:lap01}, \cite{mwpakull-E3:sno01}) that in several cases turned 
out to be {\it variable} on short time scales. We note that the optical 
selection criteria of SNR were entirely based on the relative 
strength of the low-excitation [S~II] and [O~I] lines with respect to the
H$_{\alpha}$ recombination (\cite{mwpakull-E3:mf97}), and that the diameters 
of the 'SNR' in question are exceedingly large.

\section{Optical Observations}
\label{mwpakull-E3_sec:obs}
We have selected a sample of ULX in nearby galaxies (less than 10 Mpc distance) 
mainly from the list of \cite*{mwpakull-E3:fou98} and also taking advantage
of the compilation of HRI positions by \cite*{mwpakull-E3:rw00}. At the
time of our observations no improved {\it Chandra\/} positions were known to us, 
so we took special care to correct positions by overlaying the HRI catalogue
with digitized sky atlas images available at the {\it Aladin\/} facility 
(\cite{mwpakull-E3:bon00}) at Observatoire 
Astronomique de Strasbourg. Since most X-ray images contain several sources that 
can be associated with forground stars or background AGN we have been able to
obtain improved source positions with $3\arcsec$ accuracy or better.
  
For obvious reasons, we have avoided ULX in edge-on galaxies or otherwise 
strongly absorbed (starburst) regions. These sources need to be tackled with infrared 
instrumentation, and we note that IR follow-up work on the ULX in M~82 has 
revealed a young compact star cluster at the position of the enigmatic 
most luminous ULX in this starburst galaxy (see \cite{mwpakull-E3:ebi01}).\\

The optical observations were carried out during several runs between 1999 and
2001 at  ESO NTT, CFHT, and using the Carelec spectrograph on the OHP 1.93m 
telescope. Of particular
importance for the project was the possibility of the ESO Multi-Mode Instrument 
(EMMI), and the MOS and OSIS instruments at CFHT to quickly switch between direct 
imaging and spectroscopy. Thus we could obtain spectral information of any 
photometrically selected target in the same night. Standard filters 
included B, R and H$_{\alpha}$ bands, which allowed to obtain colors and net 
emission-line images. Reduction of the data and the production of color images
was done using the MIDAS image processing software.

\begin{table}[bht]
  \caption{Some of the ULX in nearby galaxies for which we have obtained (interesting)
 optical observations.
The designations of individual sources refer to the respective discovery papers.
The following abbreviations are used: {\it neb.} = diffuse H${\alpha}$ 
emission centered on X-ray source; {\it bub.} = bubble-like nebula; 
H\,II = source within larger H${\alpha}$ emission complex; *
= possible stellar counterpart detected}  
  \label{mwpakull-E3_tab:tab1}
  \begin{center}
    \leavevmode
%    \footnotesize
    \begin{tabular}[h]{rrl}
      \hline \\[-5pt]
      Galaxy & ULX    &  nature   \\[+5pt]
      \hline \\[-5pt]
      NGC~1313  & X-1; X-2 (X-3) & neb; bub.; (SN) \\
      IC~342  & X-1; X-2 & neb.; ? \\
      Ho~II  & X-1 & XIN \\
      Ho~IX  & X-1 & bub. * \\
      M~81   & X-6 & neb. \\
      IC~2574 & X-1 & H{\sc ii} * \\
      NGC~4559 & X-1; X-7 & ?; H{\sc ii} * \\
      NGC~4631 & H7 & H{\sc ii}\\
      NGC~4861 & X-1; X-2 & H{\sc ii}; H{\sc ii}\\ 
      NGC~5204 & X-1 & bub. * \\
      NGC 7793 & P13 & neb. \\      
\hline \\
      \end{tabular}
  \end{center}
\end{table}

\vspace*{-1cm}
\section{Nebul\ae\ around ULX}
\label{mwpakull-E3_sec:nebula}
In this section we present some of the newly detected nebul\ae, 
report on their optical spectra (if available), and comment on their 
possible nature. The images show con\-ti\-nu\-um (R band) subtracted 
H$_{\alpha}$ emission in yellow, B and R band images in blue and red, 
respectively. North is up, and East is to the left.   

\subsection{IC~342 X-1}
\label{mwpakull-E3_sec:ic342}

This nearby spiral (d=4.5Mpc) harbours two variable non-nuclear 
ULX that 
\cite*{mwpakull-E3:kub01} have shown to follow soft/hard X-ray spectral 
transitions that are believed to be characteristic for black hole 
candidates. 

\begin{figure}
\centering
\includegraphics[width=\columnwidth]{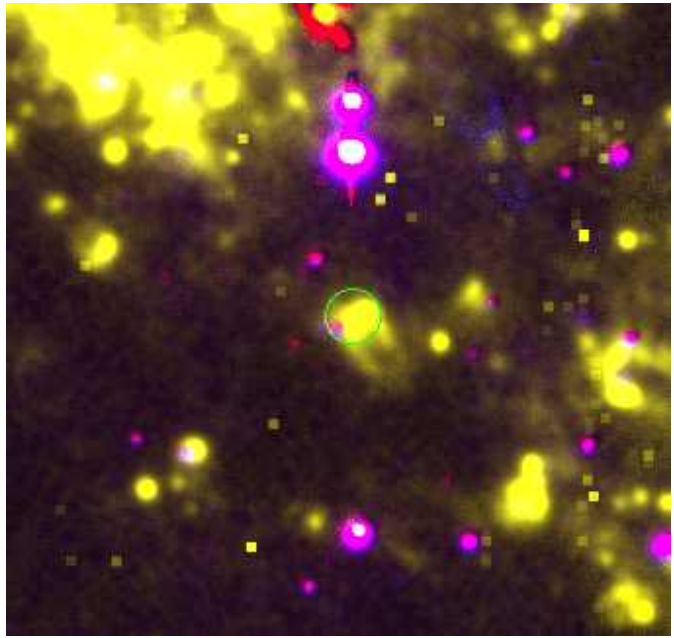}
%\vspace*{1cm}
%\includegraphics[width=8cm]{mwpakull-E3lmcx1.ps}
\caption{The area of the ULX IC~342 X-1. The error circe includes the
brightest part of the Tooth nebula which has strong low-excitation 
forbidden emission lines.}
\label{mwpakull-E3_fig:fig2}
\end{figure}

Positional information comes from the {\it ROSAT\/} HRI observations 
by \cite*{mwpakull-E3:bct93}. Figure \ref{mwpakull-E3_fig:fig2} shows
at the position of IC~342 X-1 a relatively isolated tooth-shaped nebula.
The Tooth nebula has a diameter of 220 pc and its spectrum shows 
extreme SNR-like emission line ratios: [S~II]/H$_{\alpha}=1.2$ and
[O~I]$\lambda$6300/H$_{\alpha}$=0.4 (see Sect.~\ref{mwpakull-E3_sec:xin}).

%\end{document}

\subsection{\object{Holmberg~IX X-1}}
\label{mwpakull-E3_sec:ho9}

This a close dwarf companion to the large spiral M~81, which explains 
that the X-ray source \object{Holmberg~ IX} X-1 is also known as 
\object{M~81 X-9}. At this position 
\cite*{mwpakull-E3:mi95} noted the presence of the shell-like 
nebula LH~9/10 and proposed a possible (multiple) SNR nature, 
although he also noted that the diameter of 250 pc appeared to be rather large. 
However, analysing archival X-ray data from 20 years, \cite*{mwpakull-E3:lap01} could show 
that the X-ray emission is highly variable with spectral changes reminiscent of 
of black hole candidates. 

\begin{figure}
\centering
\includegraphics[width=\columnwidth]{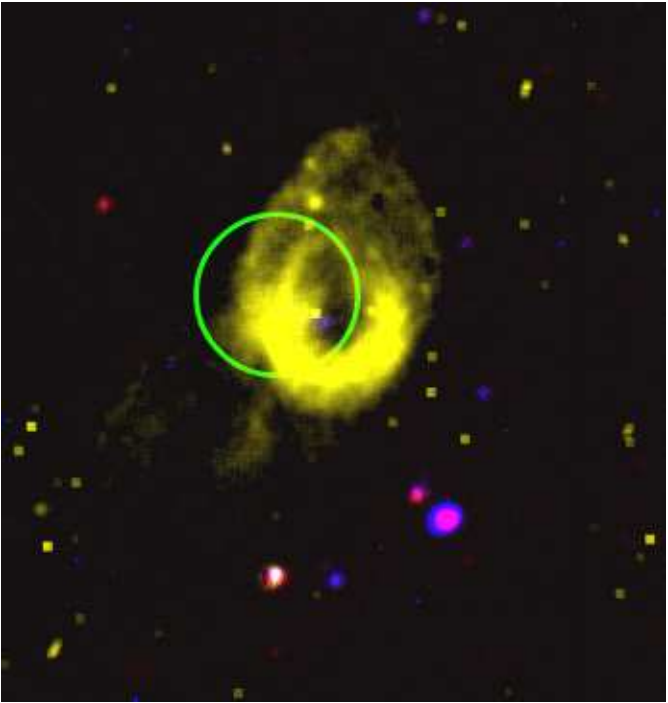}
%\vspace*{1cm}
%\includegraphics[width=8cm]{mwpakull-E3lmcx1.ps}
\caption{The barrel shaped nebula LH~9/10 surrounding 
M~81 X-9 in the dwarf galaxy companion \object{Holmberg IX}. The HRI error circle
includes a small group of faint blue stellar objects which are 
likely to be associated with the variable X-ray source.} 
\label{mwpakull-E3_fig:fig3}
\end{figure}

Figure~\ref{mwpakull-E3_fig:fig3}
shows the H$_{\alpha}$ emitting nebula LH~9/10 to have a
strikingly similar morphology to the radio images of 
barrel-shaped bilateral SNR (\cite{mwpakull-E3:gaen98}).
The brightest object in the small group of blue stars noted by 
\cite*{mwpakull-E3:mi95} has B$\sim$22.1~mag and could be a early type 
main sequence star. Our spectra confirm the high [S~II]/H$_{\alpha}$
and [O~I]$\lambda$6300/H$_{\alpha}$ ratios reported  by 
\cite*{mwpakull-E3:mi95}, but also show a more normal
Balmer decrement H$_{\alpha}$/H$_{\beta}$ rather than the 
curious value $\sim$1.6 reported in that paper. 

\subsection{\object{NGC~5204 X-1}}
\label{mwpakull-E3_sec:n5204}

\begin{figure}
\centering
\includegraphics[width=\columnwidth]{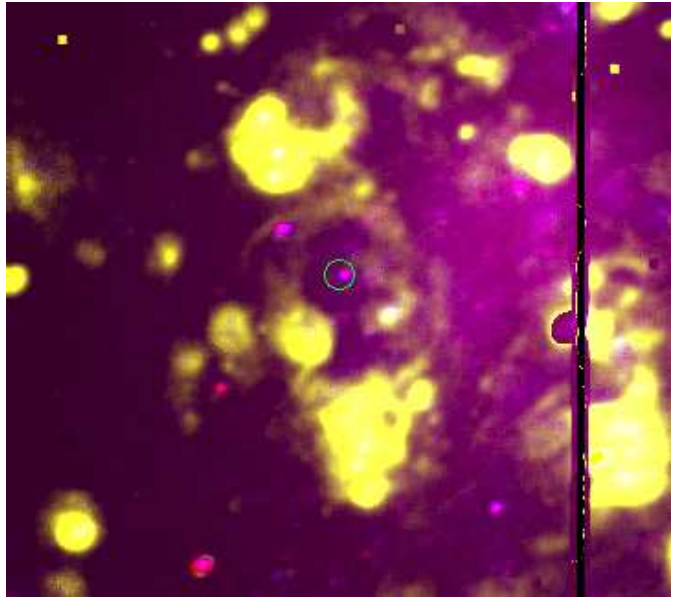}
%\vspace*{1cm}
%\includegraphics[width=8cm]{mwpakull-E3lmcx1.ps}
\caption{The area around the ULX in \object{NGC~5204} contains many  
H~{\sc ii} regions (one of them being the galactic nucleus) and 
a striking ring-like structure centered on the X-ray source.
The position obtained from {\it ROSAT\/} HRI data perfectly
matches the more precise {\it Chandra\/} position. 
%reported by 
% \cite*{mwpakull-E3:ro01}.
}
\label{mwpakull-E3_fig:fig4}
\end{figure}

\cite*{mwpakull-E3:ro01} have recently described {\it Chandra\/} and 
%{\it HST\/} 
ground based optical 
observations of the ULX in the Magellanic type galaxy \object{NGC~5204}. 
They proposed a comparatively bright V=19.7~mag stellar object to
be the optical counterpart, but on the basis of an {\it HST\/} image the
same authors (\cite{mwpakull-E3:ro02}) now find several more, fainter 
stellar images which could also be associated with the X-ray source.
Our multicolor images shown in Figure~\ref{mwpakull-E3_fig:fig4}
reveal that the B=21.9 mag star located on the eastern rim of the error
circle has very blue colors and should thus be considered a prime candidate.   

Perhaps more interesting is our discovery of a H${\alpha}$ emitting
bubble with a diameter of some 360 pc. The ring is much
larger then the "cavity" in the ionized gas reported by 
\cite*{mwpakull-E3:ro01}. Note that the field of view of their
field spectrograph was slightly smaller than the diameter of the bubble.

\subsection{NGC~4559 X-7, X-10 and IC~2574}
\label{mwpakull-E3_sec:n4559}

NGC~4559 is another galaxy in common with the work of \cite*{mwpakull-E3:ro02} 
reported at this conference. A previous {\it ROSAT\/} PSPC study was published
by \cite*{mwpakull-E3:vpb97}, and we use their source designation here.
Source X-7 is NGC~4559 X-1 in \cite*{mwpakull-E3:ro02} and is X-10 the central
object. 

Our analysis of unpublished {\it ROSAT\/} HRI
observations including positional adjustements described in section
\ref{mwpakull-E3_sec:obs} clearly shows that X-10 is {\it not} coincident
with the nucleus, providing yet another nearby ULX for further study.
X-7 is located in an isolated H~{\sc ii} region complex in the outskirts
of the galaxy, although not coincident with any of the bright nebula.
Possible counterparts are three faint blue stars at the position of the 
X-ray source. 

The nebular/X-ray source morphology is strikingly 
similar to the giant H~{\sc ii} region complex in the M\,81 group 
dwarf galaxy IC~2574, where the strong X-ray emission had been ascribed to 
extended hot plasma in an expanding supergiant shell (\cite{mwpakull-E3:walt98}).
However, inspection of the public {\it Chandra\/} image reveals that the emission
is pointlike, and our images reveal that the source is coincident with one of 
the stars in the OB association.

\subsection{NGC~1313 X-1 and X-2}
\label{mwpakull-E3_sec:n1313}

Since the pioneering {\it Einstein\/} observations of \object{NGC~1313} 
and other nearby galaxies (\cite{mwpakull-E3:ft87}) this nearby 
(4.5 Mpc distance) southern spiral has attracted much attention 
from X-ray observers. 
It harbours three ULX (\cite{mwpakull-E3:cpsr93}, \cite{mwpakull-E3:cm99}, 
\cite{mwpakull-E3:spcm00}), one of them (X-3) is the recent peculiar
supernova \object{SN~1978K}. 
The brightest source (X-1) is close to, but nevertheless clearly 
detached from the optical nucleus, and finally the outlying source X-2 
lies at a projected distance of 9 kpc from the nucleus and far from
current star formation. For some reason it
was once considered to be one of the most promising candidates of an Galactic 
cooling neutron star (\cite{mwpakull-E3:stoc95}). 

\begin{figure}
\centering
\includegraphics[width=\columnwidth,angle=270,clip=true]{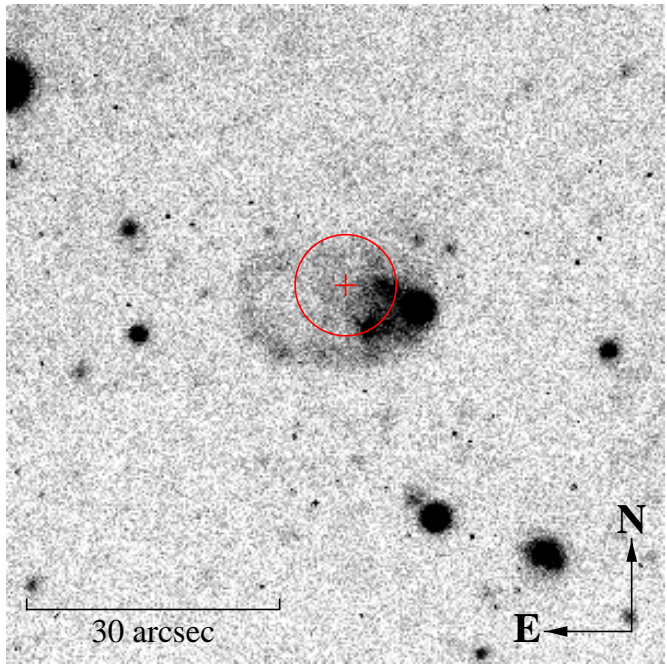}
%\vspace*{1cm}
%\includegraphics[width=8cm]{mwpakull-E3lmcx1.ps}
\caption{H$_{\alpha}$ image of the poorly populated 
neighborhood of \object{NGC~1313 X-2}.
The HRI error circle of the variable ULX is centered on the center 
of the bubble nebula and marginally excludes a relatively bright stellar object 
at the western edge. }
\label{mwpakull-E3_fig:fig5}
\end{figure}

Here we report that both NGC~1313 X-1 and X-2 imprint clear {\it optical} 
signatures of ULX activity on the interstellar medium in this galaxy.
Beginning with X-2, Fig.~\ref{mwpakull-E3_fig:fig5} shows our NTT H$_{\alpha}$ 
image on which a beautiful elongated 
bubble nebula with a large diameter of some 400pc can be seen. 
The radial velocity confirms that it is indeed located in \object{NGC~1313}.
A low resolution spectrum of the brightest region shows both relatively strong 
[S~II] and [O~I] lines; at higher resolution we were able to resolve
the emission lines to have a full-width-half-maximum width (FWHM) of 80 km/s
indicating expansion of the bubble with about the same velocity.

Perhaps less spectacular at first sight is the H$_{\alpha}$ nebula around 
\object{NGC~1313 X-1} (not shown here) which has a diameter of $\sim$240 pc.
However, our long-slit spectra reveal a well defined zone of $\sim$800 pc (!) extent 
showing an exceptionally high 
[O~I]~$\lambda$6300/H$_{\alpha}$ ratio ($>$0.1) indicating the presence 
of warm (T$_{\rm e}=10^4$~K), weakly ionized gas. This points to a situation 
where X-ray photoionization is important (see Sect.~\ref{mwpakull-E3_sec:xin}).

\subsection{Holmberg II X-1}
\label{mwpakull-E3_sec:ho2}

An even more outstanding demonstation of X-ray ionization is furnished by the
interstellar environment of the L$_{\rm x}\sim 10^{40}$ erg/s X-ray source 
in the M~81 group dwarf galaxy Holmberg~II. 
As for the other ULX cases
%(see Sect.~\ref{mwpakull-E3_sec:intro}) 
the high 
luminosity is derived from the observed flux presuming galaxian membership, 
and furthermore assuming that the source emits in an isotropic fashion.

To the best of our knowledge this source has first been discussed in the
thesis works of \cite*{mwpakull-E3:d95} and of \cite*{mwpakull-E3:fou98}.
A more recent account can be found in \cite*{mwpakull-E3:zgw99}. All
authors agree that the {\it ROSAT\/} PSPC spectrum can satisfactorily be described 
by a power-law model (photon index $\Gamma$=2.63), but from a recent {\it ASCA\/} 
observation {\cite*{mwpakull-E3:mlh01} favoured a less steep spectrum ($\Gamma$=1.89),
or a thermal spectrum with kT=4.8 keV. \cite*{mwpakull-E3:fou98} reported
that a bremsstrahlung model with kT=0.90$\pm$0.15 was an even better 
representation of the PSPC data, and she also pointed out the likely association  
with the H~{\sc ii} region HSK~70 (\cite{mwpakull-E3:hsk94}); an important finding 
that has independently been noted by \cite*{mwpakull-E3:zgw99}. 
The association is illustrated in Fig.~\ref{mwpakull-E3_fig:fig6} where we 
show the {\it ROSAT\/} error circle superposed on our CFHT multicolor image.

\begin{figure}[!ht]
\centering
\includegraphics[width=\columnwidth,angle=90,clip=true]{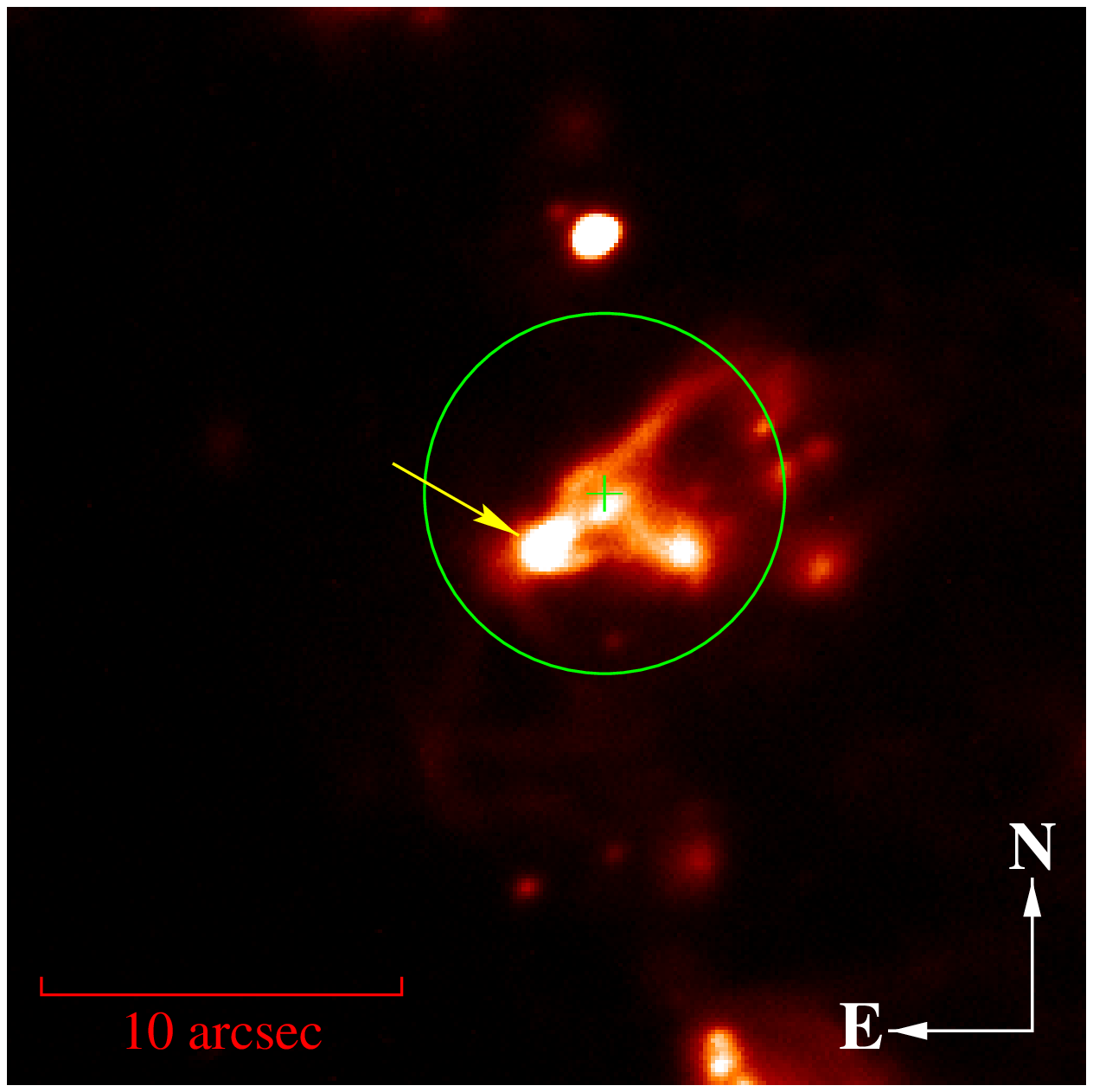}
%\vspace*{1cm}
\includegraphics[height=\columnwidth,angle=270,clip=true]{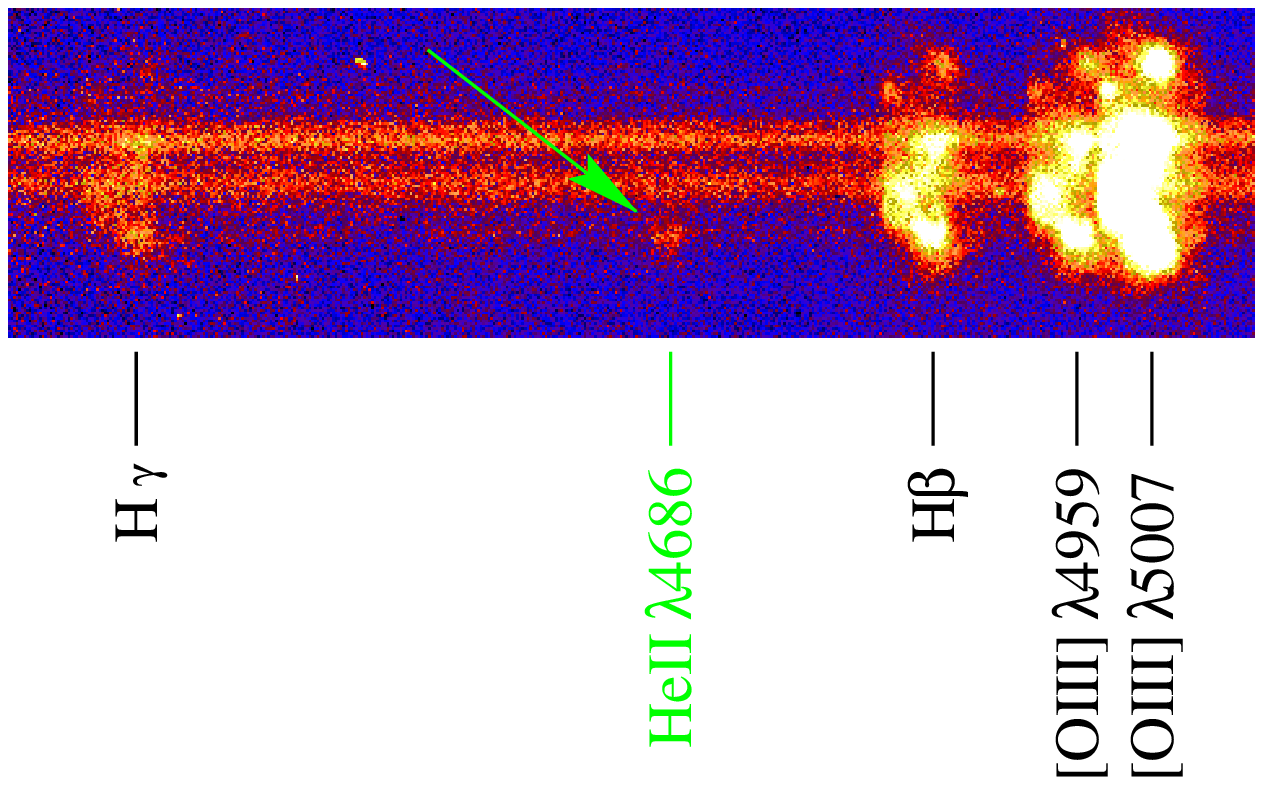}
\caption{H$_{\alpha}$ image (upper) of the H~II region HSK~70 in the dwarf 
galaxy Holmberg II with
the 5\arcsec\ radius HRI error circle superimposed. The shape resembles that of an 
X-Rayed Foot where
the emission is strongest in the Heel. The lower image is a 5\arcsec\ wide
broad-slit spectrum (i.e. essentially slit-less; the dispersion is in the N-S 
direction and the spectral resolution is essentially determined by the seeing 
of 0\farcs8) which provides monochromatic images of the narrow
nebular emission lines in the area. Note the presence of strong 
HeII$\lambda$4686 emission which is confined to the Heel.} 
\label{mwpakull-E3_fig:fig6}
\end{figure}

As the observations of
\cite*{mwpakull-E3:fou98} strongly suggested the presence of nebular 
high-excitation He~II$\lambda$4686 emission in HSK~70 we obtained a large-slit 
(slitwidth 6\arcsec)
spectrum displayed in the lower image of Fig.~\ref{mwpakull-E3_fig:fig6}, and 
which clearly shows that the $\lambda$4686 emission is confined to the 'Heel' of 
the 'Foot Nebula', and that it has has an intrinsic FWHM 
of 2\farcs2 cor\-res\-pon\-ding to 34 pc at the distance (3.2 Mpc) of Holmberg~II 
(\cite{mwpakull-E3:pmfm02}). Its luminosity L$_{\lambda4686}$ is 2.5\ 10$^{36}$ erg/s.
A narrow slit spectrum (Fig.~\ref{mwpakull-E3_fig:fig7}) of the Heel taken under
more moderate seeing conditions(FWHM=2\farcs0) is that of a
high-excitation H~{\sc ii} region typical for low metallicity
(Z/Z$_{\sun}\-\sim$0.1) starforming regions, except for the strengths of the
[O~I]\-$\lambda$6300 and particularly $\lambda$4686 emission which are truely outstanding.\

\begin{figure}[!ht]
\centering
\includegraphics[width=\columnwidth,angle=180,clip=true]{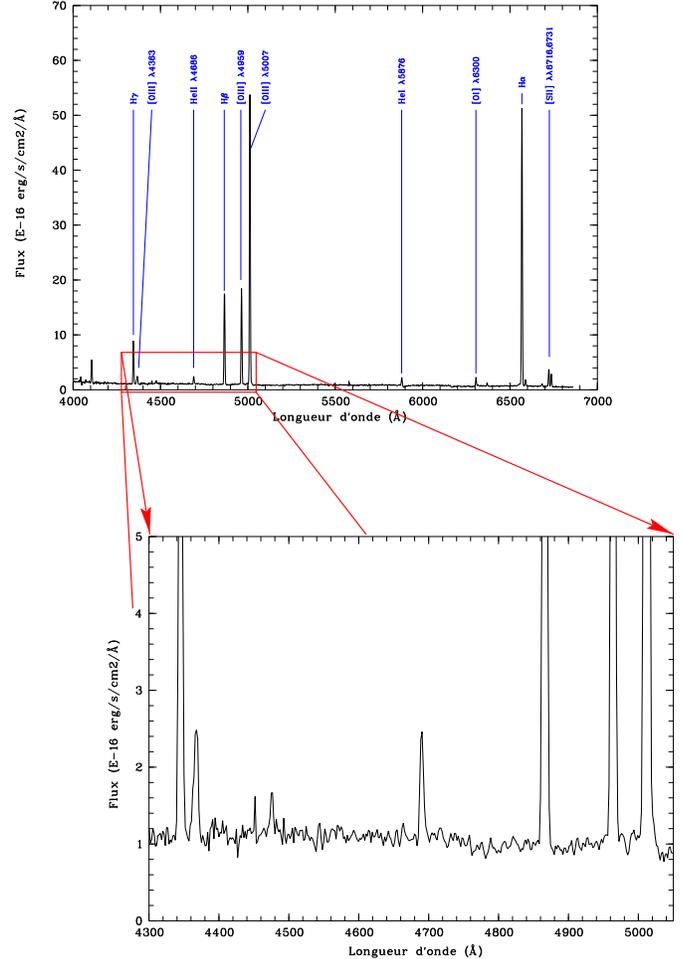}
%\vspace*{1cm}
%\includegraphics[width=8cm]{mwpakull-E3lmcx1.ps}
\caption{Narrow-slit optical spectrum of the Heel taken at OHP. The HeII$\lambda$4686
lines reaches 0.16 of the H$_{\beta}$ flux. Also note the unusually strong forbidden
neutral oxygen line [O~I]$\lambda$6300/H$_{\alpha}$ $\sim$0.03.
 } 
\label{mwpakull-E3_fig:fig7}
\end{figure}

Its luminosity L$_{\lambda4686}$ is 2.5\ 10$^{36}$ erg/s.
A narrow slit spectrum (Fig.~\ref{mwpakull-E3_fig:fig7}) of the Heel taken under
more moderate seeing conditions(FWHM=2\farcs0) is that of a
high-excitation H~{\sc ii} region typical for low metallicity
(Z/Z$_{\sun}\-\sim$0.1) starforming regions, except for the strengths of the
[O~I]\-$\lambda$6300 and particularly $\lambda$4686 emission which are truely outstanding.\\

We therefore propose that we are observing the optical manifestation
of an X-ray ionized nebula (XIN), most probably somewhat diluted by more conventional
ionization by O stars. The radial velocity agrees with that of the other nearby 
H~{\sc ii} regions in Holmberg~II and proves that the X-ray source is 
indeed situated {\it within} that galaxy. Compared to XIN N159F around the black hole
candidate LMC X-1 (c.f. Sect.~\ref{mwpakull-E3_sec:prosp}) Holmberg~II X-1 
is $\sim$30$\times$ more luminous, both in X-rays and in He~II$\lambda$4686 emission. 

Finally, in order illustrate the structure of the XIN we display in 
Fig.~\ref{mwpakull-E3_fig:fig8} various emission line 
intensities and the continum as a function of position along a 1\farcs5 wide 
long-slit orientated E-W and centered on the Heel of the Foot. We see that the maximum
of the $\lambda$4686 emission more or less coincides with the peak of the Balmer
emission and with a non-resolved, B=20.6~mag continuum source. 
However, closer inspection reveals that the He~II$\lambda$4686/H$_{\beta}$ ratio 
(not shown here) clearly peaks towards 
the East of the line maxima. On the other hand, the [O~I]$\lambda$6300 intensity
distribution is clearly shifted towards the "Toes". In 
Section~\ref{mwpakull-E3_sec:xin} we will argue that this is just what one would 
expect for the case that the nebula is density bounded in the eastern direction.

\begin{figure}
\centering
\includegraphics[width=\columnwidth]{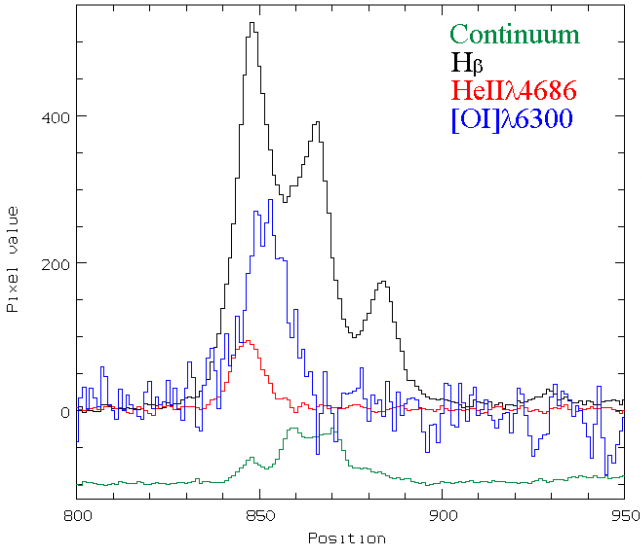}
%\vspace*{1cm}
%\includegraphics[width=8cm]{mwpakull-E3lmcx1.ps}
\caption{Intensity distribution of of various emission lines and the continuum 
as a function of pixel position (1~pix=0\farcs28) of a long-slit orientated 
East (left) - West (right).
Relative intensities are not drawn to scale. Note in particular the relative
shift between the [O~I]$\lambda$6300 and He~II$\lambda$4686 emission regions, where
the indicating the presence of warm O$^0$ towards the 'Toes'.} 
\label{mwpakull-E3_fig:fig8}
\end{figure}

\section{Remnants, winds and XIN}
\label{mwpakull-E3_sec:xin}

The discovery of nebula around a significant fraction of ULX undoubtedly provides
clues to their formation and to their mass-loss history, either though explosive
events or by stellar winds or jets. ULX NGC~1313 X-2 will
serve as an example for what can be deduced from the optical observations.
Assuming that the 400 pc diameter nebula is the remnant of a supernova-like event
which created the compact star we can use the well-known relations of SNR evolution.
The large extent clearly suggests that the remnant must be in the pressure driven
snowplough phase, where the evolution of the outer shock proceeds by
(\cite{mwpakull-E3:cmb88})
\begin{equation}
{\rm R} = 93\ {\rm pc}\ E_{52}^{0.22}\ n^{-0.257}\ t_{6}^{0.3}\\
\end{equation}
\begin{equation}
{\rm v} = 28\ {\rm km/s}\ E_{52}^{0.22}\ n^{-0.257}\ t_{6}^{-0.7}
\end{equation}
where R is the radius and v the expansion velocity of the remnant after
10$^6$ t$_{6}$ yrs, E$_{52}$ the explosion energy in units of 10$^{52}$ erg and 
and n is the interstellar density (cm$^{-3}$) into which the remnant expands.
If both R and v can be measured, we can solve for E 
\begin{equation}
{\rm E} = 6.8\, 10^{43}\ {\rm erg}\ R_{\rm pc}^{3.16}\ v_{\rm km/s}^{1.36}\ n^{1.16}
\end{equation}
The density can estimated using the \cite*{mwpakull-E3:ds96} 
scaling relations of the total radiative H$_{\beta}$ flux 
(including radiative precursor) emitted by interstellar shocks of 
100 v$_{100}$ km/s velocity
\begin{equation}
{\rm F}_{\beta} = 1.7\ 10^{-5}\ v_{100}^{2.35}\ {\rm n}\ {\rm erg}\ {\rm cm}^{-2}\ {\rm s}^{-1}
\end{equation}
Plugging in the observed values for the NGC~1313 X-2 nebula 
(R=200pc; v=80 km/s; L$_{\beta}$ $\sim$ 1$\times$10$^{37}$ erg/s) 
we arrive at an age of $\sim$0.8 Myr, density $\sim$0.2 cm$^{-3}$ 
and an impressive explosion energy of $\sim$ 3-10$\times$ 10$^{52}$ erg/s.

Alternatively, we could imagine that the bubbles are being inflated by
ongoing stellar wind or jet activity, such as observed in 
\object{SS~433} (c.f. Sect.~\ref{mwpakull-E3_sec:prosp}). Although in this
case the numerical factors and exponents in Eq. (1)-(3) change somewhat,
the total energy requirements and active lifetimes would be about the same.
The implied wind/jet luminosity ($=0.5\dot{M}\ v_{jet}^2$) 
of 1.5 10$^{39}$ erg/s can thus only be generated in a {\it relativistic} outflow,  
such as from the \object{SS~433} system.\\

We now turn to the physics of X-ray ionized nebul\ae\,, and to what we can
learn about the sources that excite them. This field was pioneered by
\cite*{mwpakull-E3:tts69}, developped further in \cite*{mwpakull-E3:kmc82},  
and applied to XIN N159F/LMC X-1 by
\cite*{mwpakull-E3:pa86}. 
The main difference to more conventional H~{\sc ii}
regions is the absence of sharp transitions between ionized and
neutral plasma at the outer boundary (Str\"omgren spheres), since X-rays
are not very efficiently absorbed. This creates an extended warm (electron temperature 
T$_{\rm e}\sim$ 10$^4$ K) weakly ionized zone, in which neutral
atoms can be collisionally excited. 

The emission from highly ionized gas like the  
He$^{++}\rightarrow\,$He$^+$ $\lambda$4686 recombination line close to the source, 
and from forbidden transitions of 
neutrals like [O~I]$\lambda$6300 in the outer extended zones are thus the 
hallmarks of XIN. Note that $\lambda$4686 acts as a photon counter of 
the emitted source flux in the He$^+$ Lyman continuum between 54 eV 
and about 200 eV (\cite{mwpakull-E3:pa86}). 
This also implies that few $\lambda$4686 photons will be emitted if the 
source is intrinsically weak or absorbed below $\sim$100 eV.

Taking advantage of the widely used photoionisation 
code {\it CLOUDY} developed by Gary \cite*{mwpakull-E3:f96}, we have calculated 
a grid of XIN with a range of 
brems\-strah\-lung input spectra (including various contributions of normal stellar 
ionizing continua), nebuar density and metallicity. These were confronted with 
our spatially resolved spectra of the Heel Nebula in Holmberg~II. 
The results can be summarized as follows (see \cite{mwpakull-E3:pmfm02}):

\begin{itemize}
 \item We confirm the photon counting property of the $\lambda$4686 line 
 irrespective of metallicity, density or presence of O star continua.  
 \item The ionizing X-ray continuum is diluted by more conventional O star
 radiation. This lowers the relative strength of $\lambda$4686 as compared to
 H$_{\beta}$, and decreases [O~I]$\lambda$6300/H$_{\alpha}$ due to 
 photionization by O stars of O$^0$ in the extended warm zone of the XIN. 
 \item The fact that the $\lambda$4686 and $\lambda$6300 emitting regions do 
 not spatially coincide practically excludes the remote possibility of strong 
 shock excitation in the nebula. The absence of [O~I]$\lambda$6300 emission 
 to the east of the heel (Fig.~\ref{mwpakull-E3_fig:fig8})
 can be understood by a density-bounded geometry in that direction.
 \item For an assumed thermal bremsstrahlung spectrum of temperature $kT$
 the observed $\lambda$4686 luminosity of 2.5\ 10$^{36}$ erg/s implies that
 the nebula actually "sees" (and reprocesses) the unabsorbed He$^+$ Lyman continuum 
 of such a spectrum having an X-ray luminosity of 
 some 3$\times$10$^{39}$\ $\times kT_{keV}$ erg/s.  
 Accordingly we deduce for the range of temperatures summarized in 
 Sect.~\ref{mwpakull-E3_sec:ho2} an expected X-ray output 
 L$_{\rm X}$ = 0.3 - 1.3\ 10$^{40}$ erg/s, which is in agreement with the
 directly observed luminosity. {\it Therefore we think that our data exclude
 significantly non-isotropic beaming for Holmberg II X-1}.  
\end{itemize}
To conclude this Section we emphasize that optical spectra of fast interstellar
shocks (e.g. \cite{mwpakull-E3:ds96}) which are possibly present in some of 
our ULX nebul\ae, do often look quite similar
to XIN. Compared to normal H~{\sc ii} regions both display enhanced forbidden lines 
of neutral species, and He~II$\lambda$4686 can become noticable for shock velocities
larger than 300 km/s. The inclusion of radiative precursors of shocks can furthermore
lower a very high electron temperature in the post shock 
O$^{++}$ region (observable from the the weel-known $\lambda$5007/$\lambda$4363
line ratio) to more conventional 1-2$\times$10$^4$ K.

\section{Conclusions}
\label{mwpakull-E3_sec:conc}

We are aware of the fact that we have just begun to scratch the surface of a new (optical)
observational approach to study the enigmatic ULX, and possibly to understand formation
and evolution of luminous compact X-ray sources in general.

We find, like other authors before us, that ULX do preferentially occur in, or close
to star forming regions, and that some nearby starbursts harbour several of them. 
Yet, the bubbles in \object{Holmberg~IX} and \object{NGC~1313} are
located far from other young objects. Maybe there are two populations of ULX ?
 
Among the more quiet galaxies, ULX appear to prefer the more metal-poor 
dwarfs. Examples are the M81 group, with only one such object in M81 itself,
but ULX are seen in IC~2574, NGC 2403, Holmberg~II and Holmberg~IX; the M101 group has
only the ULX in NGC~5204... \ It is well-known that massive stellar evolution proceeds
differently in such environments (\cite{mwpakull-E3:maed92}), mainly because wind
loss is much reduced in low-metallicity stars during the pre-SN evolution. 
This naturally leads to much more massive cores that acoordingly might collapse 
to more massive black holes.

\begin{acknowledgements}
We thank Gary Ferland for providing the photoionisation code {\it CLOUDY} and
the Brief Introduction {\it Hazy}. We also acknowledge frequent use of {\it SIMBAD},
{\it VizieR} and {\it Aladin} operated at Centre de Donn\'ees
astronomiques de Strasbourg. 
\end{acknowledgements}

\end{document}